\documentclass[prl,twocolumn,amsmath,nofootinbib,amssymb,superscriptaddress,floatfix]{revtex4}

\usepackage{graphicx}
\usepackage{pdfsync}
\usepackage{epsfig,psfrag,amsmath,amssymb,float}

\begin{document}

\newcommand{\ltwid}{\mathrel{\raise.3ex\hbox{$<$\kern-.75em\lower1ex\hbox{$\sim$}}}} 
\newcommand{\gtwid}{\mathrel{\raise.3ex\hbox{$>$\kern-.75em\lower1ex\hbox{$\sim$}}}} 
\def\K{{\bf{K}}} 
\def\Q{{\bf{Q}}} 
\def\Gbar{\bar{G}} 
\def\tk{\tilde{\bf{k}}} 
\def\k{{\bf{k}}}
\def\n{\langle n\rangle}

\title{Dynamic cluster quantum Monte Carlo simulations of a two-dimensional
Hubbard model with stripe-like charge density wave modulations: Interplay between
inhomogeneity and superconductivity}

\author{T.A.~Maier} \affiliation{Computer Science and Mathematics Division and
Center for Nanophase Materials Sciences,\\ Oak Ridge National Laboratory, Oak
Ridge, TN 37831-6164} \email{maierta@ornl.gov}

\author{G.~Alvarez} \affiliation{Computer Science and Mathematics Division and
Center for Nanophase Materials Sciences,\\ Oak Ridge National Laboratory, Oak
Ridge, TN 37831-6164} \email{alvarezcampg@ornl.gov}

\author{M.~Summers} \affiliation{Computer Science and Mathematics Division and
Center for Nanophase Materials Sciences,\\ Oak Ridge National Laboratory, Oak
Ridge, TN 37831-6164} \email{summersms@ornl.gov}

\author{T.C.~Schulthess} \affiliation{Institut f\"ur Theoretische Physik, ETH
Z\"urich, 8093 Z\"urich, Switzerland} \email{schulthess@cscs.ch}

\date{\today} 

\begin{abstract} Using dynamic cluster quantum Monte Carlo simulations, we
study the superconducting behavior of a 1/8 doped two-dimensional Hubbard model
with imposed uni-directional stripe-like charge density wave modulation. We find a
significant increase of the pairing correlations and critical temperature
relative to the homogeneous system when the modulation length-scale is
sufficiently large. With a separable form of the irreducible particle-particle
vertex, we show that optimized superconductivity is obtained for moderate
modulation strength due to a delicate balance between the modulation enhanced
pairing interaction, and a concomitant suppression of the bare
particle-particle excitations by a modulation reduction of the quasi-particle
weight.  \end{abstract}

\maketitle

Despite decades of intense research, there is currently no general consensus on
a theory of the pairing mechanism in the high-temperature superconducting
cuprates.  This is in part due to the many complex phenomena observed in the
cuprates, and the lack of insight and agreement as to which are or are not
relevant for superconductivity. For example, nano-scale charge and spin
inhomogeneities \cite{tranquada:nature95}, as well as random gap modulations
\cite{lang:nature02,hanaguri:nature04,gomes_visualizing_2007} have been found
to emerge in a number of cuprates. These observations raise several interesting
questions regarding the interplay between inhomogeneities and superconductivity
\cite{emergy:physicac93,Kivelson:2005p941}: Do inhomogeneities cause
high-temperature superconductivity or are they merely spectators? Do they
enhance or suppress the pairing mechanism? And perhaps most importantly from an
application point of view, is there an optimum inhomogeneity that maximizes the
transition temperature? 

ARPES \cite{Valla:2006p4260} and transport measurements
\cite{Tranquada:2008p1043} indicate that superconductivity is optimized in some
respect in the striped state in LaBaCuO.  While true 3-dimensional
superconductivity is absent, the transport experiments show the existence of
2-dimensional superconductivity over a wide temperature range suggesting
coexistence between stripe and superconducting phases. On the theory side,
studies of the interplay between inhomogeneity and superconductivity have been
largely phenomenological or used toy models
\cite{Arrigoni:2002p936,Arrigoni:2003p940,Martin:2005p944,Aryanpour:2006p4299,
Aryanpour:2007p1521,Tsai:2006p942,Loh:2007p1673,Mishra:2008p3571,Yao:2007p1704},
with only a few recent exceptions \cite{tsai_optimal_2008,Doluweera:2008p1642}.
Estimates of $T_c$ in large enough systems with repulsive interactions and a
realistic representation of the inhomogeneity are still lacking. 

Here, we study the effect of charge stripes, realized as a uni-directional
charge density wave modulation, on the pairing correlations and $T_c$ in a
two-dimensional (2D) Hubbard model. To this end, we impose a charge modulation
by applying a spatially varying local potential $V_i$, and then study its
impact on superconductivity. We note that $V_i$ is phenomenological and as such
has no direct microscopic origin that corresponds to a degree of freedom in the
actual materials. This approach is justified when the characteristic energy
scale for the formation of charge stripes exceeds the superconducting energy
scale, i.e. the gap energy $\Delta$. Specifically, we are interested in the
question of whether one can have a higher transition temperature in a striped
array in which there is strong pairing in the spin-correlated low hole density
regions, and good hole mobility in the higher hole density regions.

Our study is based on a Hubbard model on a square 2D 
lattice with near neighbor hopping $t$ and Coulomb repulsion $U$,
given by
\begin{eqnarray}
\label{HM}
{H} &=& -t\sum_{\langle ij\rangle,\sigma}
(c^\dagger_{i\sigma}c^{\phantom\dagger}_{j\sigma}+ {\rm h.c.}) + U\sum_i
n_{i\uparrow}n_{i\downarrow}\nonumber\\
 &-& \sum_{i\sigma}(\mu+V_i) n_{i\sigma}\,,
\end{eqnarray}
%\begin{equation} \label{HM} {H} = -t\sum_{\langle ij\rangle,\sigma}
%(c^\dagger_{i\sigma}c^{\phantom\dagger}_{j\sigma}+ {\rm h.c.}) + U\sum_i
%n_{i\uparrow}n_{i\downarrow} - \sum_{i\sigma}(\mu+V_i) n_{i\sigma}\,,
%\end{equation}
with $\mu$ the chemical potential which sets the filling $\langle n\rangle$.
Here $c^\dagger_{i\sigma}$ creates an electron with spin $\sigma$ on site $i$
and $n_{i\sigma}=c^\dagger_{i\sigma}c^{\phantom\dagger}_{i\sigma}$ is the site
occupation operator for spin $\sigma$. 

To study the effects of the charge density modulation, we have used a
dynamic cluster approximation (DCA) \cite{hettler:dca1,Maier:2005p682} with a
Hirsch-Fye quantum Monte Carlo (QMC) cluster solver \cite{Maier:2005p682}.
Large cluster simulations with this technique find a superconducting transition
at finite temperature in the 2D homogeneous Hubbard model
\cite{maier_systematic_2005-3}, and have allowed us to reveal the nature of the
pairing interaction responsible for it
\cite{maier_structure_2006-1,maier:prb06,maier_dynamics_2008-1}. This approach
therefore provides an ideal framework for the present study of the
inhomogeneous model. The general idea of the DCA is to approximate the effects
of correlations in the bulk lattice with those on a finite size cluster with
$N_c$ sites and periodic boundary conditions. The DCA maps the bulk lattice
problem onto an effective periodic cluster embedded in a self-consistent
dynamic mean-field that is designed to represent the remaining degrees of
freedom. We have used an $L_x$$\times$$L_y$-site cluster with $L_x=8$ and
$L_y=4$. This cluster is large enough to accomodate the experimental situation
of 1/8 doped LaBaCuO, where neutron scattering finds a periodicity of 4 lattice
spacings in the charge sector, and 8 lattice spacings in the magnetic channel
\cite{tranquada:nature95}. We point out, however, that our simulation takes
place in the paramagnetic phase without spin order. We adjust the
chemical potential $\mu$ so that the average filling $\n=0.875$, and
set the local Coulomb repulsion to $U=4$ in units of the hopping $t$. For the
32-site cluster, this allows simulations at low temperatures with a manageable
QMC fermion sign problem.
\begin{figure}[htbp]
\includegraphics[width=3.25in]{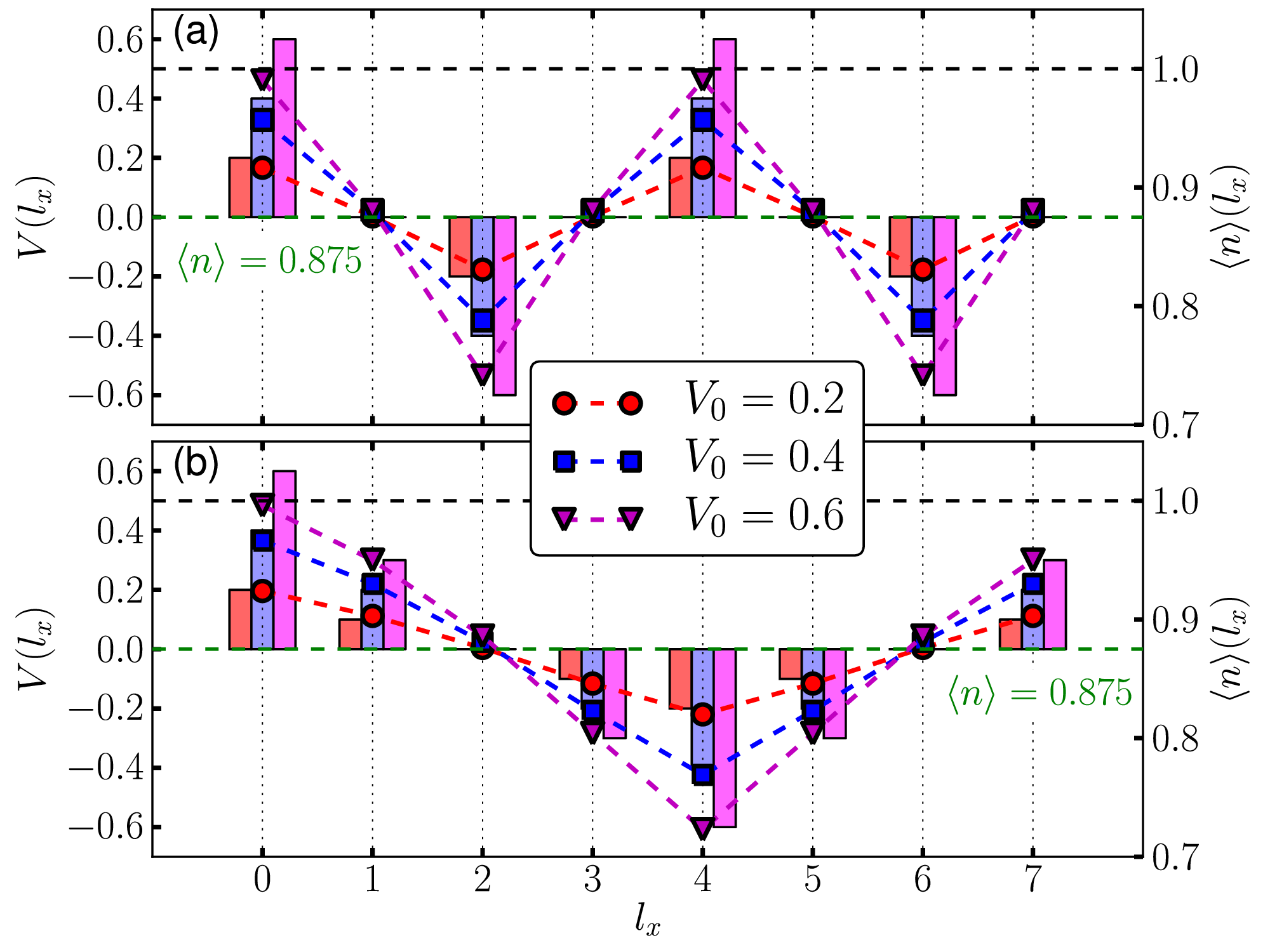} \caption{(Color online) Charge
modulation potential $V(l_x)$ (bars) and resulting local occupation $\langle n
\rangle(l_x)$ (symbols connected by dashed lines) versus location $l_x$ along
the long side in the 8$\times$4-site cluster. The system has translational
invariance along $L_y$. Results are for a temperature $T=0.125$ and modulation
wave-vector (a) $Q=\pi/2$ and (b) $Q=\pi/4$. The dashed green line indicates
the average filling of $\langle n \rangle=0.875$.}
\label{fig1}
\end{figure}

In order to impose a smooth charge modulation along $L_x$ with translational
invariance along $L_y$, we have chosen a smoothly varying potential $V(l_x)$
with magnitude $V_0$ and modulation wave-vector $Q$, as shown in
Fig.~\ref{fig1} as bars. To study the effect of different modulation length
scales, we consider two cases with $Q=\pi/2$ (Fig.~\ref{fig1}a) and $Q=\pi/4$
(Fig.~\ref{fig1}b).  The resulting site occupation along $L_x$, calculated at a
temperature $T=0.125$ is also shown in the figure for different magnitudes
$V_0$ of the potential. As one can see, the variation of the site occupation
follows closely the variation of the potential $V(l_x)$.

In order to keep the problem computationally tractable, we average the cluster
results over different stripe locations along $L_x$, i.e.  different phases of
the modulation potential, before the mean-field medium is computed.  This
corresponds to a situation where the stripe order is short-ranged, over the
length-scale of the cluster, but the system has translational invariance on
longer macroscopic length-scales. After averaging, translational invariance is
restored and the off-diagonal components of the single-particle cluster Green's
function $G(\K,\K')$ where $\K$ and $\K'$ are cluster wave-vectors, and
similarly for the two-particle correlation functions vanish.

In order to determine the pairing correlations and critical temperature, we
compute the eigenvalues and eigenvectors of the Bethe-Salpether equation in the
particle-particle channel \cite{maier_structure_2006-1}
\begin{eqnarray}
\label{BSE}
-\frac{T}{N_c}\sum_{K'} \Gamma^{pp}(K,K')\bar{\chi}_0^{pp}(K')\phi_\alpha(K') = 
\lambda_\alpha\phi_\alpha(K)\,.
\end{eqnarray}
Here, $K=(\K,i\omega_n)$ and $\Gamma^{pp}(K,K')$ is the irreducible
particle-particle vertex with center of mass momentum $Q=0$ calculated on the
cluster. The coarse-grained bare particle-particle Green's function
$\bar{\chi}_0^{pp}(K')=N_c/N\sum_{{\tilde k}'} G_\uparrow(\K'+\tk',\omega_n)
G_\downarrow(-\K'-\tk',-\omega_n)$ is calculated from the lattice Green's
function $G(\k',\omega_n) =
[i\omega_n-\epsilon_{\k'}+\mu-\Sigma(\K',\omega_n)]^{-1}$ with the dispersion
$\epsilon_{\k'} = -2t(\cos k'_x + \cos k'_y)$ and the cluster self-energy
$\Sigma(\K',\omega_n)$.  When the leading eigenvalue $\lambda_\alpha$ becomes
one, the system undergoes a superconducting transition, and the symmetry of the
corresponding state is determined by the wave-vector dependence (and frequency
dependence) of the corresponding eigenvector $\phi_\alpha(K)$. In all the cases
we have studied, we find that the leading eigenvalue occurs in the spin
singlet, even frequency channel, and the corresponding eigenvector has
dominantly $d_{x^2-y^2}$ symmetry.
\begin{figure}[b]
\includegraphics[width=3.25in]{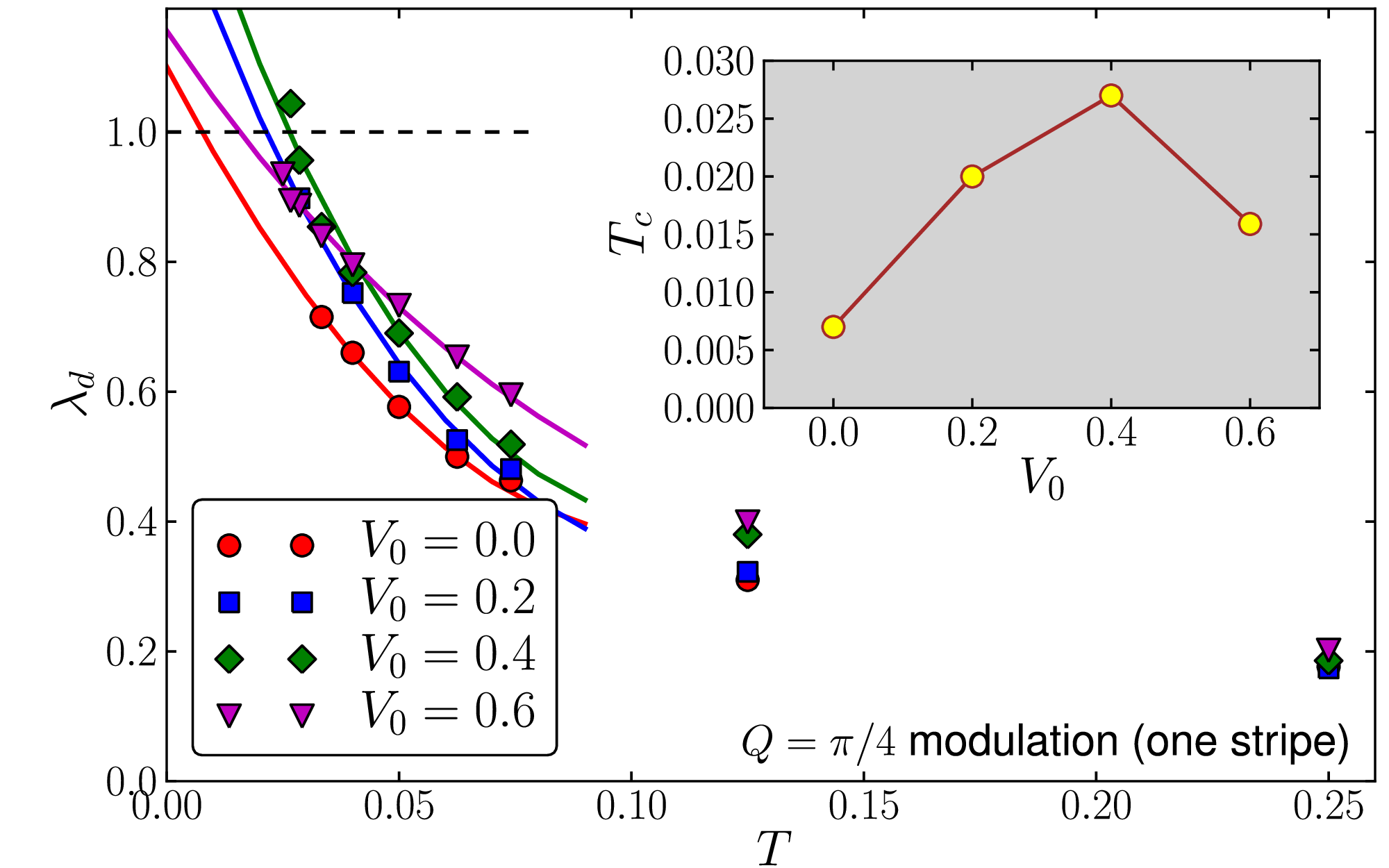} \caption{(Color online) The leading
($d$-wave) eigenvalue of the particle-particle Bethe-Salpeter equation for the
homogeneous and  inhomogeneous systems with a modulation periodicity of 8 sites
($Q=\pi/4$) for different modulation strengths $V_0$. The stripe inhomogeneity
significantly enhances the pairing correlations and $T_c$. The transition
temperature $T_c$ as a function of modulation strength $V_0$ is shown in the
inset.}
\label{fig3}
\end{figure}

We start by discussing the results for the system with the $Q=\pi/4$ modulation
(one stripe). The temperature dependence of the leading eigenvalue
$\lambda_d$ for this case is shown in Fig.~\ref{fig3}. As one can see,
the inhomogeneity significantly enhances the pairing correlations as
indicated by the size of the eigenvalue $\lambda_d$, as well as the
critical temperature $T_c$ as given by the temperature where
$\lambda_d$ crosses one.  For moderate temperatures ($T\sim 0.05 -
0.12$), $\lambda_d$ increases monotonically with modulation strength
$V_0$. At lower temperatures ($T<0.05$), however, $\lambda_d$ for
$V_0=0.6$, while still enhanced over the $V_0=0$ result, drops below
the results for $V_0=0.4$ and even $V_0=0.2$. 

In order to estimate the critical temperature $T_c$, we inter- and extrapolate
the low temperature results for $\lambda_d$ as a function of temperature. For
the inhomogeneous cases, we were able to perform simulations down to
temperatures where $\lambda_d$ is already larger than one, or very close to
one. The resulting estimates for $T_c$ are therefore reliable. For the
homogeneous system with $V_0=0$, the fermion sign problem prevents us
from reaching temperatures below $T\sim 0.03$. The corresponding
estimate for $T_c$ is therefore less reliable, but it is clear from
the results that $T_c$ for $V_0=0$ is significantly smaller than $T_c$
for finite $V_0$.  The inset to Fig.~\ref{fig3} shows the resulting
estimates of $T_c$ plotted versus the modulation strength $V_0$. As
one can see, $T_c$ is optimized for moderate modulation strength
$V_0=0.4$. For $V_0=0.6$, $T_c$ is reduced from the critical
temperature for $V_0=0.4$, due to the flattening of $\lambda_d(T)$ at
low temperatures.
\begin{figure}[b]
\includegraphics[width=3.5in]{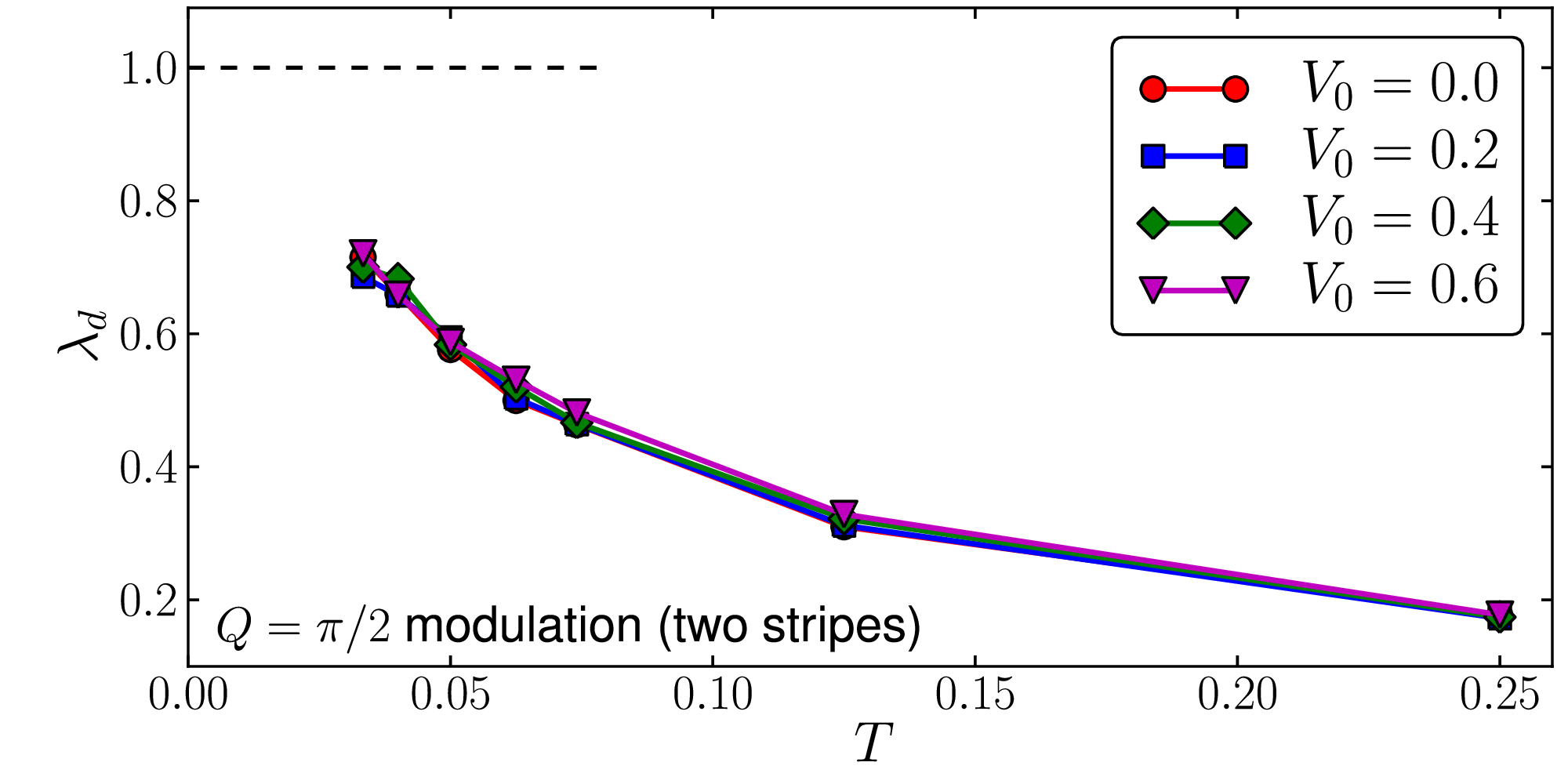} \caption{(Color online) The leading
($d$-wave) eigenvalue of the particle-particle Bethe-Salpeter equation for the
homogeneous system ($V_0=0$) and the inhomogeneous systems (finite $V_0$) with
a modulation periodicity of 4 sites ($Q=\pi/2$). The shorter length-scale
inhomogeneity has no effect on the pairing correlations.}
\label{fig2}
\end{figure}

The temperature dependence of the leading eigenvalues $\lambda_d$ for
the system with the shorter length-scale $Q=\pi/2$ modulation
corresponding to two stripes in the 8$\times$4 cluster is shown in
Fig.~\ref{fig2}.  As one can see, the eigenvalues of the inhomogeneous
systems are essentially identical to the eigenvalue of the homogeneous
system, indicating that the modulation with $Q=\pi/2$ has no effect on
the pairing correlations and the critical temperature $T_c$. 

Our results are consistent with the work by Martin {\it et al.}
\cite{Martin:2005p944}, where an attractive Hubbard model with a
similar modulation of the attractive interaction $U$ and also the
charge was studied in a mean-field BCS treatment. There, the authors
found that $T_c$ was unaffected by the inhomogeneity, if the
modulation length-scale was small compared with the coherence length.
Conversely, an enhancement of $T_c$ was found when the modulation
length-scale was of the same order as the coherence length.  We do not
have a direct estimate of the coherence length or the size of the
Cooper pairs in our calculation. However, a natural explanation of the
strong increase of the pairing correlations for $Q=\pi/4$ as seen in
Fig.~\ref{fig3} and their insensitivity towards the $Q=\pi/2$
modulation shown in Fig.~\ref{fig2}, is that in the latter case the
inhomogeneity is simply averaged out, because the modulation
length-scale is short compared with the size of the Cooper pairs.

Our results show that the experimentally relevant period 4 stripes do not
enhance the pairing instability, while a longer wave-length modulation leads to
a strong enhancement. This is in contrast to recent experiments which indicate
that the period 4 charge modulation is optimum for the LaBaCuO material
\cite{Tranquada:2008p1043}. A logical reason for this discrepancy could be that
the ``coherence length'' or Cooper pair size in our simulations is large
compared to the actual materials, perhaps due to the relatively small Coulomb
repulsion $U=4$ we have used. A systematic study of this issue is reserved for
future work.

To gain more insight into the inhomogeneity induced enhancement of the pairing
correlations for $Q=\pi/4$, we have constructed a separable
representation of the pairing interaction $\Gamma(K,K')$,
\begin{eqnarray}
\label{eq:sep} \Gamma(K,K') = -V_d\phi_d(K)\phi_d(K')
\end{eqnarray}
with the leading $d$-wave eigenvector $\phi_d(K)$
\cite{maier:prb06}. With this separable form and Eq.~(\ref{BSE}),
one finds that
\begin{eqnarray}
\label{eq:Vd}
V_d\frac{T}{N_c}\sum_{K'}\phi_d^2(K') \bar{\chi}_0^{pp}(K') = \lambda_d\,,
\end{eqnarray}
allowing us to determine a strength $V_d$ of the separable interaction from
Eq.~(\ref{eq:sep}). The strength of $V_d$ depends upon both the site
occupation $\langle n\rangle$ and the temperature. In
Ref.~\cite{maier:prb06} $V_d$ was found to increase with
decreasing temperature and increasing site occupation in a homogeneous
2D Hubbard model. But although $V_d$ increases as $\n$ goes to one,
the number of holes available for pairing, as measured by the quantity 
\begin{eqnarray} \label{eq:Pd0} 
P_d^0(T) = \frac{T}{N_c}\sum_K \phi_d^2(K)\bar{\chi}_0^{pp}(K)\,, 
\end{eqnarray} 
was found to be suppressed with increasing $\n$ due to the vicinity to
the Mott state where the quasiparticle weight goes to zero. Because of
this, $T_c$ decreases to zero as $\n$ goes to one.  In a system with a
spatial modulation of $\n$, one would therefore expect a corresponding
modulation of $V_d$ and $P_d^0$. If the system can take advantage of
the strong pairing interaction $V_d$ in the spin correlated regions
with low hole-density, and of the increased $P_d^0$ in the hole-rich
regions, the pairing correlations and $T_c$ could be enhanced.
\begin{figure}[h]
\includegraphics[width=3.25in]{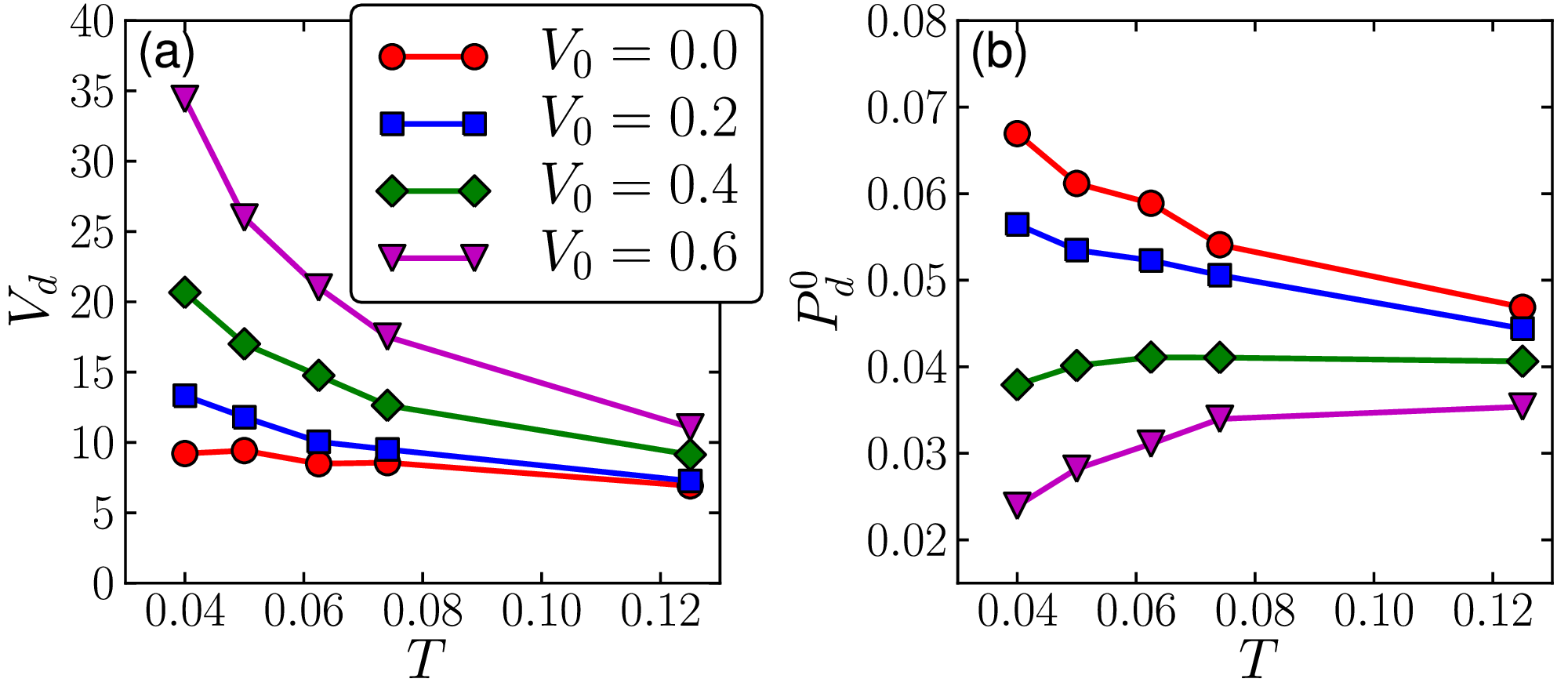} \caption{(Color online) The pairing
interaction $V_d(T)$(a) defined in Eq.~(\ref{eq:Vd}) and the bare d-wave
pairing susceptibility $P_d^0(T)$ defined in Eq.~(\ref{eq:Pd0}) as a function of
temperature for different modulation strength $V_0$ for the modulation with
$Q=\pi/4$. The pairing interaction $V_d$ is enhanced by the charge modulation,
while $P_d^0$ is reduced.}
\label{fig4}
\end{figure}

For the system with the $Q=\pi/4$ charge modulation, Fig.~\ref{fig4} shows the
results for $V_d(T)$ in (a) and $P_d^0(T)$ in (b) versus temperature for
different modulation strength $V_0$, obtained after the averaging over stripe
locations has been performed. One clearly sees that the pairing interaction
$V_d$ increases with modulation strength $V_0$, while the hole mobility $P_d^0$
shows the opposite behavior. Thus there is a delicate balance between the two
effects, which can either increase or decrease the pairing correlations:
%For the homogeneous system with $V_0=0$, we find that the increase in
%$\lambda_d(T)$ arises mainly from the increase in $P_d^0(T)$, as $V_d(T)$ is
%approximately constant over the low temperature region.  Interestingly, the
%reverse behavior is found for the system with optimized superconductivity
%($V_0=0.4$): $P_d^0(T)$ is flat, but $V_d(T)$ strongly increases as the
%temperature is lowered. In this
For the system with optimal inhomogeneity ($V_0=0.4$), the increase in
$\lambda_d(T)$ and $T_c$ arises, because the enhancement of $V_d$
overcompensates the reduction of $P_d^0$.  For stronger modulation ($V_0=0.6$),
a larger reduction of $P_d^0$ tips the balance towards a reduction of
$\lambda_d$ at low temperature relative to the system with $V_0=0.4$ (see
Fig.~\ref{fig3}). As can be seen from Fig.~\ref{fig1}, the hole density in the
regions between the charge stripes is almost zero ($\n=1$) for $V_0=0.6$. Thus,
while it is beneficial to have regions with strong antiferromagnetic
correlations in between the hole-rich regions, it is also favorable to have
finite hole density remaining in the spin-correlated regions.

In summary, we have studied the superconducting behavior of a 2D
inhomogeneous Hubbard model with imposed uni-directional 
charge density wave modulation with wavelengths $Q=\pi/2$ and $\pi/4$ using a
dynamic cluster quantum Monte Carlo approximation for an
8$\times$4-site cluster. We find a significant increase of the pairing
correlations and $T_c$ in the system with the long $Q=\pi/4$
modulation length scale. Optimized superconductivity with the highest
$T_c$ is obtained for moderate modulation strength, due to a delicate
balance between the inhomogeneity enhanced pairing interaction, and a
concomitant suppression of the bare particle-particle excitations by a
modulation reduction of the quasi-particle weight. While experiments
indicate optimized pairing for a $Q=\pi/2$ modulation, we find that in
this case inhomogeneity has no effect on the pairing correlations. A
possible explanation for this discrepancy is a difference in coherence
lengths between our simulation and the real materials.

\acknowledgements We would like to acknowledge useful discussions with
D.J.~Scalapino and S.~Okamoto. This research was enabled by
computational resources of the Center for Computational Sciences at
Oak Ridge National Laboratory and conducted at the Center for
Nanophase Materials Sciences, which is sponsored at Oak Ridge National
Laboratory by the Division of Scientific User Facilities, U.S.
Department of Energy.

\bibliography{bib}

\end{document}